\documentclass[
,showpacs
]{revtex4}

\begin{document}
\title{Coating the carbon nanotubes: Geometry of incommensurate long-range ordered 
physisorbed monolayers}
\author{Antonio \v{S}iber}
\email{asiber@ifs.hr}
\affiliation{Institute of Physics, P.O. Box 304, 10001 Zagreb, Croatia}
\begin{abstract}
The structures of long-range ordered physisorbed monolayer on a carbon nanotube are 
examined. Geometrical and energetical constraints determining the order of 
such monolayers are discussed. A number of symmetrically different, strongly bound 
adsorbate structures is found for Xe adsorbates, some of which differ very little in 
energy. The presented results suggest that the atomically uniform coating of 
carbon nanotubes is possible and offer a clear visualization of such 
coatings.
\end{abstract}
\maketitle
\begin{center}
{\bf Published in Phys. Rev. B 68, 033406 (2003)}
\end{center}

Carbon nanotubes have already found a number of interesting applications \cite{HDai}. 
However, to truly assess their potential for various 
applications, a detailed description of interaction of carbon nanotubes 
with atoms and molecules of their surroundings is required. In this context, 
it has been suggested that coating of carbon nanotubes with atoms may 
improve their properties that are important for various uses (see e.g. Refs. 
\onlinecite{Ajayan1} and \onlinecite{Ajayancoat}). The experimental demonstration 
of coating of 
carbon nanotubes with crystalline sheets of V$_2$O$_5$ layer-like structures 
has been presented in Ref. \onlinecite{Ajayancoat}. However, it is not 
clear yet how to create fine, ordered coatings of monoatomic thickness. 
The long-wavelength vibrational properties of such coatings have been 
examined in the literature (see e.g. Ref. \onlinecite{Vidales}), 
but the coatings are not easily visualized since the confinement 
of atoms/molecules 
on a nanotube (cylindrical) surface imposes a specific 
ordering of such structures, as will be shown here.

In this letter, we are interested in geometry and binding 
energies of long-range 
ordered monoatomically thick coatings of carbon nanotubes. We consider 
the capped nanotubes and thus, the monolayer coating is on the 
outer side of the nanotube (exohedral adsorption). The microscopic 
arrangement of the adsorbed layer of atoms depends on the details of 
both the adsorbate-adsorbate and adsorbate-substrate interactions. These 
interactions are not precisely known yet with the exception of physisorbed 
gases. A large amount of experimental 
and theoretical data exists for physisorption of gases on 
graphite surfaces \cite{Bruchbook}, which enables one to use this 
information and reconstruct the relevant potentials that can 
then be applied to the present case of interest \cite{uptake,SibBulj,sibphon,sibgroove}. 
Even with the information on potentials, it is still not trivial 
to find the ordered adsorbate structures that minimize the free 
energy of the system. This is mainly because there are many different configurations 
with very similar energy. The adsorbate-substrate potential 
exhibits corrugation and in the case of very large corrugation amplitudes, 
the adsorbate overlayer will follow the periodicity of the nanotube, 
since the energy it looses due to the unfavorable positions of 
its neighbors is less than the one gained due to the fact that 
all the atoms sit in the minima of the substrate potential, i.e. 
the overlayer structure will be commensurate with the substrate. In the other 
extreme case in which the corrugation of the substrate is negligible, the adsorbate 
layer will order in such a way to maximize the binding energy resulting 
from the interadsorbate interactions, and the resulting structure 
will be incommensurate \cite{Back}. The corrugation of graphite 
is in between the two extremes, and a number of different phases 
of gases physisorbed on the graphite surface (commensurate and incommensurate) 
has been reported in the literature \cite{Bruchbook}, depending also on the substrate 
temperature. In what follows, I shall assume that the 
adsorbate structure is indeed incommensurate with the corrugation of 
the carbon nanotube. In that case, the adsorbate-nanotube potential 
can be treated as smooth, i.e. one can neglect its corrugation when 
calculating the total energy of the structure.

Assume for the moment that all of the adsorbate atoms are at the same distance 
from the nanotube axis, i.e. that their centers lie on a cylinder of radius 
$R$, $R=R_t+h$, where $R_t$ is the nanotube radius and $h$ is the height of 
the coating above the surface of the nanotube. 
It can be {\em a priori} expected (I shall not do so) that the 
adsorbate structure with minimum energy will locally have a nearly hexagonal order. 
It is nevertheless not completely obvious how to reconcile 
this local arrangement of atoms with the fact that {\em all} adsorbate 
sites on a cylinder surface must be equivalent.

We define two 
vectors (${\bf a}_1$ and ${\bf a}_2$) on a cylinder surface 
connecting nearest neighboring adsorbate atoms along 
two inequivalent directions. Any of the two vectors can be specified by 
a polar angle spanned by the vector's starting and ending points, and 
by the projection of the vector on the cylinder axis ($z$ axis),
\begin{eqnarray}
{\bf a}_1 &=& (\alpha, a \alpha), \nonumber \\
{\bf a}_2 &=& (\beta, b \beta).
\label{eq:bvc}
\end{eqnarray}
The vectors are defined on a cylinder surface i.e. each of them specifies 
one of the cylinder geodesics which are simple coil curves 
(see Fig. \ref{fig:fig1}). The lengths of 
the vectors are not necessarily the same and it is in principle 
possible that the nearest neighbor distances are not the same along the 
two geodesics.

In order to obtain 
a long range ordered adsorbate structure, all the adsorbate position vectors  
${\bf r}_{l}$ should be linear combinations of the 
elementary vectors ${\bf a}_1$ and ${\bf a}_2$, i.e.
\begin{eqnarray}
{\bf r}_l &\equiv & {\bf r}_{i,j} = i {\bf a}_1 + j {\bf a}_2 \nonumber \\
&=& (R \cos (i \alpha + j \beta), R \sin (i \alpha + j \beta), 
i a \alpha + j b \beta),
\label{eq:rij}
\end{eqnarray}
where $i$ and $j$ are integers, and in the last line of the above equation 
${\bf r}_{i,j}$ are given in Cartesian coordinates. 
There is an additional condition that must be satisfied by the parameters 
of the structure. Starting from one crossing point of two geodesics, the 
next crossing point should be reached by integer numbers $(m,n)$  
of primitive vector translations along both geodesics. This yields
\begin{eqnarray}
m \alpha &=& n \beta + 2 \pi , \nonumber \\
m a \alpha &=& n b \beta .
\label{eq:deter}
\end{eqnarray}
The adsorbate structure is thus defined by $\alpha, \beta, a$, and 
$b$ which satisfy Eq. (\ref{eq:deter}).

The total energy of the adsorbate structure can be written as a sum of 
all adsorbate-adsorbate interactions ,$v$, and interaction of the adsorbates 
with the carbon nanotube, $V$ (this neglects the energy of adsorbate zero-point 
motion). The energy per adsorbate atom is thus
\begin{equation}
E_b = { \sum_{i=-\infty}^{\infty} \sum_{j=0}^{n-1} } \, ' \; v({\bf r}_{i,j} 
- {\bf r}_{0,0}) + V({\bf r}_{0,0}),
\label{eq:energ}
\end{equation}
where ${\bf r}_{i,j}$ is given by Eq. (\ref{eq:rij}), and the prime on the sum 
indicates that the term with $i=j=0$ is excluded. 

We now look for the structure with minimum adsorbate energy, i.e. we minimize 
Eq. (\ref{eq:energ}) with respect to its parameters. Explicitly, 
the minimization procedure consists in choosing the height of the overlayer $h$, 
$\alpha$ and $a$ parameters of the first geodesic, integer numbers 
$m$ and $n$ that determine $\beta$ and $b$ of the 
second geodesic [Eq.(\ref{eq:deter})], 
and calculating the energy of the resulting structure from 
Eq. (\ref{eq:energ}). Since we minimize the internal energy of the 
overlayer and not its free energy, our results 
are applicable at zero temperature only.

The indexing of the structures with $(m,n)$ is not unique because the integers 
$m$ and $n$ depend on the two nearest-neighbor geodesics one chooses (e.g. 
instead of two geodesics defined by ${\bf a}_1$ and ${\bf a}_2$ one could 
have chosen the two with $-{\bf a}_1$ and ${\bf a}_2$, or
${\bf a}_1$ and ${\bf a}_1 \pm {\bf a}_2$, presuming that 
${\bf a}_1 \pm {\bf a}_2$ still points to the nearest neighbor, see 
Fig. \ref{fig:fig1}). However, the 
representation of adsorbate structures of interest in terms of planar 
adsorbate arrangements folded onto a cylinder surface is 
possible and allows for a unique indexing. All strongly bound structures of interest 
exhibit quasi-hexagonal order of adsorbates, with all six nearest neighbor atoms 
nearly equally separated from the atom in the center of a hexagon. Nevertheless, 
these quasi-hexagons cannot in general be obtained by 
mathematical wrapping of the {\em perfect} hexagonal planar lattice of minimal energy.
This is a consequence of the fact that the sum of 
inter-adsorbate interactions is different when the adsorbates are positioned 
on the surface of a cylinder (Cartesian distances change), and that the whole 
structure must "fit" onto the cylinder. To specify the geometry of adsorbate 
coating it is sufficient 
to characterize two points in the planar, nearly hexagonal structure 
obtained by unwrapping the coating (the cutting preceding the unwrapping 
is parallel to the $z$-axis). The vector ${\bf w}$ 
connecting these two points can be thought of as being wrapped around the $z=0$ 
cross-section of the cylinder. The whole planar structure is assumed to follow this 
"wrapping" procedure. This is quite similar to the usual way of indexing of 
carbon nanotubes \cite{HDai}. The wrapping procedure is sketched in Fig. \ref{fig:fig1}.

The vector ${\bf w}$ can also be expressed as
\begin{equation}
{\bf w} = m {\bf a}_1' - n {\bf a}_2',
\label{eq:wv}
\end{equation}
where ${\bf a}_1'$ and ${\bf a}_2'$ are basis vectors of the planar, nearly 
hexagonal adsorbate structure which, when wrapped around the cylinder, produce 
the vectors ${\bf a}_1$ and ${\bf a}_2$ in Eq. (\ref{eq:bvc}), and the integers 
$m$ and $n$ correspond to those in Eq. (\ref{eq:deter}). If one chooses 
${\bf a}_1'$ and ${\bf a}_2'$ in the way sketched in Fig. \ref{fig:fig1} 
(angle between ${\bf a}_1'$ and ${\bf a}_2'$ greater than $\pi /2$), the 
indexing of coatings in terms of $m$ and $n$ [or 
wrapping vector in Eq. (\ref{eq:wv})] becomes unique and 
that is what we shall use to specify the symmetries of the coatings. Note that the 
minus sign in Eq. (\ref{eq:wv}) is a convention: one can also 
choose the plus sign and redefine vectors ${\bf a}_1'$ and ${\bf a}_2'$, so that 
the angle between them is smaller than $\pi /2$. Alternatively, $m$ and $n$ 
numbers are those pertaining to ${\bf a}_1$ and ${\bf a}_2$ vectors chosen in 
such a way that the angle they span is maximum, while 
$\alpha a \beta b > 0$ [Eq.(\ref{eq:bvc})].

We now investigate the symmetries and binding energies of the 
adsorbate coatings as a function of the carbon nanotube radius. 
Adsorption of xenon on nanotube bundles has been 
experimentally studied (see Refs. \onlinecite{Migone,Kuznec}) 
and molecular dynamics simulations of Xe 
adsorption on carbon nanotube have led the authors of Ref. \onlinecite{Simonyan} 
to conclude that the corrugation of the Xe-nanotube potential is not large 
enough to impose commensurate Xe structure. That is why the Xe monolayer coatings
are considered in the following, 
although the approach presented here can be easily extended to other adsorbates which 
physisorb on nanotubes. 

The Xe-Xe ($v$) and Xe-nanotube ($V$) potentials are constructed 
as described in Ref. \onlinecite{uptake} (see also Refs. \onlinecite{sibphon,cvitas}). 
These are not the most precise potentials, 
but are fairly simple and shall suffice to illustrate the most important points 
of this article.

\begin{table*}[t]
\begin{ruledtabular}
\begin{tabular}{|c|c|c|c|c|c|c|c|}
\multicolumn{4}{c}{$R_t$=3.39 \AA; (5,5) nanotube} &
\multicolumn{4}{c}{$R_t$=6.78 \AA; (10,10) nanotube} \\
\hline
Configuration & ${\bf w}$; $(m,n)$ & $h$ [\AA] & $E_b$ [meV] & Configuration & 
${\bf w}$; $(m,n)$ & $h$ [\AA] & 
$E_b$ [meV] \\
\hline
0 & (4,7) & 3.710 & -231.54 & 0 & (4,12) & 3.730 & -242.23 \\
\hline
1 & (1,9) & 3.705 & -231.41 & 1 & (1,14) & 3.740 & -242.10 \\
\hline
2 & (5,6) & 3.700 & -231.29 & 2 & (5,11) & 3.720 & -241.88 \\
\hline
3 & (3,8) & 3.740 & -231.06 & 3 & (3,13) & 3.720 & -241.72 \\
\hline

\end{tabular}
\end{ruledtabular}
\caption{Summary of the key characteristics of four lowest energy coatings 
with different symmetries for 
(5,5) and (10,10) single-walled carbon nanotubes. Wrapping vector (${\bf w}$), height 
above the carbon nanotube surface ($h$) and binding energy ($E_b$) are specified.}
\label{tab:tab1}
\end{table*}

Table 1 summarizes the results of calculations of 
binding energies per adsorbate atom for several lowest energy configurations 
of Xe coatings and for two different radii of a single wall carbon nanotube. The 
configurations are numbered as 0,1,2,... and are ordered such 
that their energy 
increases with the increase of the configuration number. The 
radii chosen correspond to (5,5) and (10,10) carbon nanotubes.
The symmetries of the coatings are denoted. The four lowest energy Xe structures 
on (5,5) carbon nanotube are plotted in Fig. \ref{fig:fig2}. The parameters 
for the lowest energy structure ($m=4$, $n=7$) on (5,5) nanotube are 
$(\alpha =0.506$ rad, $\beta=-0.609$ rad, $a =-5.650$ \AA /rad, $b=2.681$ \AA /rad), while 
for (4,12) coating on (10,10) nanotube, $(\alpha =0.302$ rad, $\beta=-0.423$ rad, 
$a=-10.87$ \AA /rad, $b=2.583$ \AA /rad).

One should bear in mind that it is possible to find a 
configuration of Xe coating with the energy in between those specified in 
Table 1. Such a structure can be realized 
by slightly straining the interadsorbate distances in the 
minimum energy configuration of a particular symmetry in such a way that the 
symmetry of the structure remains the same. For example, 
the minimum energy structure of a particular symmetry can be turned into 
the structure of the same symmetry but smaller binding energy, by applying 
forces to the coating along the $z$ direction to slightly 
squeeze (or stretch) it. The discrete energies quoted in Table 1 correspond to the lowest 
energy configurations of a coating with specified symmetry and it is not possible 
to obtain lower energy by applying any set of {\em infinitesimal} changes to the 
adsorbate position vectors.
Obviously, a large number of different symmetry coatings differing 
only a little in the binding energy is obtained. For example, for (10,10) nanotube 
highly bound coatings are also those with wrapping indices (2,13), (6,10), 
(0,14), (8,9), (7,9) ... But the coatings with 
different symmetries do not all find the minimum energy configuration 
at the same distance from the tube center. This is due to the fact that 
it may be advantageous to the 
overlayer to slightly move away from the height at which it interacts 
with the nanotube most attractively, in order to "repack" itself in 
such a way to increase a part of the binding energy which is due to 
interadsorbate interactions. All the energies are in fact {\em doubly 
degenerate}, since for each choice of ($\alpha, \beta, a, b$), there is 
a structure of exactly the same energy, but topologically different, 
with parameters ($\alpha, \beta, -a, -b$). This 
could be termed as helicity degeneracy. Futhermore, any rotation of the 
coating around the $z$-axis does not change its energy. 

Note also that {\em the adsorbate coatings do not in general show 
periodicity in $z$-direction}. This is due to the fact that the 
unwrapped 2D hexagonal lattice is not perfect (see Fig. \ref{fig:fig1}). 
Only when $m \alpha = 2 f \pi$, where $f$ is a 
rational number, and $m$ and $\alpha$ those from Eq. (\ref{eq:deter}), the 
adsorbate structure is periodic in $z$-direction. 

For multiwalled carbon nanotubes, the conclusions reached at here will not 
change significantly. In that case only the adsorbate-substrate potential change, 
and the procedure described above can applies equally well. Nevertheless, 
due to small energy differences between different configurations of the coatings, it is 
possible that the minimum energy configurations may exhibit different 
symmetries for single-walled nanotube and multiwalled nanotube with the same 
(external) radii. The presented approach can be also applied to the coatings which 
form {\em inside} the nanotubes. Essentially the only difference in this 
case is that $h$ is negative. The packing of fullerene molecules in 
carbon nanotubes has been considered in 
Ref. \onlinecite{Girifalco}. Packing of hard spheres in cylinders has 
been studied in Ref. \onlinecite{Picket}. The authors of this reference 
emphasize the chirality aspect of the structures which is also obtained in this work. 
The lattices of hard spheres formed on cylindrical surfaces have been studied in 
Ref. \onlinecite{Ericks}.

There is an additional degree of freedom available to the coating which was 
effectively "frozen" in the presented approach. Namely, one could imagine a 
situation where the adsorbate monolayer additionally relaxes in such a way 
that its height above the nanotube becomes positional dependent. This is similar 
to the phenomenon of the overlayer "buckling" encountered in some 
adsorbate/planar surface systems of for some clean 
surfaces (see e.g. Ref. \onlinecite{buckl}). The simplest form of "buckled" 
coating would be the one with the atomic centers positioned on an 
ellipse in the $x-y$ plane (instead of a circle). 

An interesting question which comes to mind, especially in view of such small 
energy differences among various configurations, is whether the inclusion of 
zero point motion energy would change the results presented in a qualitative way. 
The different structures are indeed expected to have inequal  
zero-point motion energies, since the potentials experienced by adsorbates  
differ in these configurations, at least slightly. This is probably 
a very small effect whose evaluation may require quite sophisticated techniques.

In conclusion, we have considered geometrical and energetical constraints 
and requirements associated 
with coating of the nanotube with a monolayer of physisorbed atoms. We have 
theoretically predicted a number of different symmetry coatings, some of which 
differ only a little in their binding energy. The results should be of 
importance for future experiments directed towards functionalization 
of carbon nanotubes by such fine, monoatomically thick coatings.

Stimulating and usefull discussions with D. Jurman, Dr. B. Gumhalter and Dr. H. Buljan 
are acknowledged.

\begin{figure}[h]
\caption{Nearly perfect hexagonal lattice of points. The way to turn this 
lattice into an adsorbate 
coating is illustrated by the dashed lines and "scissors" (cutting along 
the dashed lines and wrapping the ${\bf w}$ vector around $z=0$ cross-section of 
the nanotube). The basis 
vectors ${\bf a}_1'$ and ${\bf a}_2'$ used to 
specify any wrapping vector ${\bf w}$ are denoted. Several minimum energy structures 
obtained 
for (5,5) nanotubes are indicated. Note that although all these structures are 
indicated on the {\em same} hexagonal lattice, they in fact belong to 
{\em slightly different} 
nearly perfect hexagonal lattices as discussed in the text. The inset displays 
${\bf a}_1$ and ${\bf a}_2$ vectors and their components.}
\label{fig:fig1}
\end{figure}

\begin{figure}[h]
\caption{Four lowest energy Xe structures on (5,5) carbon nanotube. From left to right: 
(4,7), (1,9), (5,6) and (3,8) coatings.}
\label{fig:fig2}
\end{figure}


\begin{thebibliography}{}
%
\bibitem{HDai} H. Dai, Surf. Sci. {\bf 500}, 218 (2002)
%
\bibitem{Ajayan1} P.M. Ajayan and T.W. Ebbesen, Rep. Prog. Phys. {\bf 60}, 
1025 (1997)
%
\bibitem{Ajayancoat} P.M. Ajayan, O. Stephan, Ph. Redlich, and 
C. Colliex, Nature {\bf 564}, 375 (1995)
%
\bibitem{Vidales} A.M. Vidales, V.H. Crespi, and M.W. Cole, Phys. Rev. B 
{\bf 58}, R13426 (1998)
%
\bibitem{Bruchbook} L.W. Bruch, M.W. Cole, and E. Zaremba, {\em Physical 
Adsorption: Forces and Phenomena}, (Clarendon Press, Oxford, 1997)
%
\bibitem{Back} P. Bak, Rep. Prog. Phys. {\bf 45}, 587 (1982)
%
\bibitem{uptake}G. Stan, M.J. Bojan, S. Curtarolo, S.M. Gatica, and M.W. Cole, 
Phys. Rev. B {\bf 62}, 2173 (2000) 
%
\bibitem{SibBulj} A. \v{S}iber and H. Buljan, Phys. Rev. B {\bf 66}, 075415 (2002)
%
\bibitem{sibphon} A. \v{S}iber, Phys. Rev. B {\bf 66}, 235414 (2002)
%
\bibitem{sibgroove} A. \v{S}iber, Phys. Rev. B {\bf 66}, 205406 (2002)
%
\bibitem{Migone} S. Talapatra, A.Z. Zambano, S.E. Weber, and A.D. Migone, 
Phys. Rev. Lett. {\bf 85}, 138 (2000); A.J. Zambano, S. Talapatra, and 
A.D. Migone, Phys. Rev. B {\bf 64}, 075415 (2001)
%
\bibitem{Kuznec} A. Kuznetsova, J.T. Yates, Jr., J. Liu, and R.E. Smalley, 
J. Chem. Phys. {\bf 112}, 9590 (2000)
%
\bibitem{Simonyan} V.V. Simonyan, J.K. Johnson, A. Kuznetsova, and J.T. Yates, Jr., 
J. Chem. Phys. {\bf 114}, 4180 (2001)
%
\bibitem{cvitas} M.T. Cvita\v{s} and A. \v{S}iber, Phys. Rev. B {\bf 67}, 193401 (2003)
%
\bibitem{Girifalco} M. Hodak and L.A. Girifalco, Phys. Rev. B {\bf 67}, 075419 (2003)
%
\bibitem{Picket} G.T. Picket, M. Gross, and H. Okuyama, Phys. Rev. Lett. 
{\bf 85}, 3652 (2000)
%
\bibitem{Ericks} R.O. Erickson, Science {\bf 181}, 705 (1973)
%
\bibitem{buckl} A. Zangwill, {\em Physics at Surfaces}, 
(Cambridge University Press, 1988)
%
\end{thebibliography}
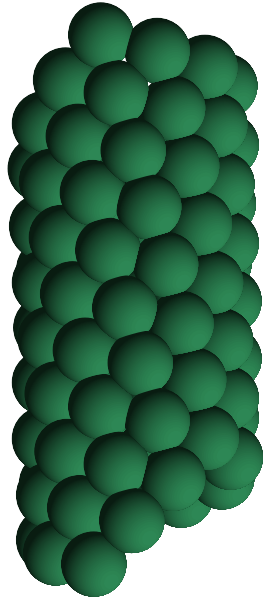
\end{document}